\documentstyle[twoside,12pt]{article}

\catcode`\@=11
\long\def\@makefntext#1{ 
\protect\noindent \hbox to 3.2pt {\hskip-.9pt
$^{{\eightrm\@thefnmark}}$\hfil}#1\hfill} 

\def\thefootnote{\fnsymbol{footnote}}
 \def\@makefnmark{\hbox to 0pt{$^{\@thefnmark}$\hss}}  

\def\ps@myheadings{\let\@mkboth\@gobbletwo
\def\@oddhead{\hbox{} 
\rightmark\hfil\eightrm\thepage}
\def\@oddfoot{}\def\@evenhead{\eightrm\thepage\hfil 
\leftmark\hbox{}}\def\@evenfoot{}
\def\sectionmark##1{}\def\subsectionmark##1{}}

\oddsidemargin=\evensidemargin
\renewcommand{\thefootnote}{\fnsymbol{footnote}}

\newcounter{sectionc}\newcounter{subsectionc}\newcounter{subsubsectionc}
\renewcommand{\section}[1] {\vspace{12pt}\addtocounter{sectionc}{1}
\setcounter{subsectionc}{0}\setcounter{subsubsectionc}{0}\noindent
        {\bf\thesectionc. #1}\par\vspace{5pt}}
\renewcommand{\subsection}[1] {\vspace{12pt}\addtocounter{subsectionc}{1}
        \setcounter{subsubsectionc}{0}\noindent
        {\bf\thesectionc.\thesubsectionc. {\kern1pt \bf\it
#1}}\par\vspace{5pt}}
\renewcommand{\subsubsection}[1] {\vspace{12pt}\addtocounter{subsubsectionc}{1}
        \noindent{\thesectionc.\thesubsectionc.\thesubsubsectionc.
        {\kern1pt \it #1}}\par\vspace{5pt}}
\newcommand{\nonumsection}[1] {\vspace{12pt}\noindent{\bf #1}
        \par\vspace{5pt}}

\newcommand{\textlineskip}{\baselineskip=14pt}

\def\eightcirc{
\begin{picture}(0,0)
\put(4.4,1.8){\circle{6.5}}
\end{picture}}
\def\eightcopyright{\eightcirc\kern2.7pt\hbox{\eightrm c}}

\def\abstracts#1#2#3{{
        \centering{\begin{minipage}{5in}\baselineskip=12pt\tenrm
        \centerline{ABSTRACT}
        \parindent=0pt #1\par
        \parindent=15pt #2\par
        \parindent=15pt #3
        \end{minipage} }\par}}

\renewenvironment{thebibliography}[1]                   
        {
         \begin{list}{\arabic{enumi}.}                  
        {\usecounter{enumi}\setlength{\parsep}{0pt}
         \setlength{\leftmargin 17pt}{\rightmargin 0pt} 
         \setlength{\itemsep}{0pt} \settowidth          
        {\labelwidth}{#1.}\sloppy}}{\end{list}} 

\newcounter{itemlistc}
\newcounter{romanlistc}
\newcounter{alphlistc}
\newcounter{arabiclistc}

\newcounter{tempfigtabc}                        

\def\pmb#1{\setbox0=\hbox{#1}
        \kern-.025em\copy0\kern-\wd0
        \kern.05em\copy0\kern-\wd0
        \kern-.025em\raise.0433em\box0}

\def\fnt#1#2{\footnotetext{\kern-.3em
        {$^{\mbox{\scriptsize #1}}$}{#2}}}

\def\fpage#1{\begingroup
\voffset=.3in
\thispagestyle{empty}\begin{table}[b]\centerline{\footnotesize #1}
        \end{table}\endgroup}

 1
 1
 1

\font\eightrm=cmr8

\newskip\humongous \humongous=0pt plus 1000pt minus 1000pt
\def\caja{\mathsurround=0pt}
\def\eqalign#1{\,\vcenter{\openup1\jot \caja
        \ialign{\strut \hfil$\displaystyle{##}$&$
        \displaystyle{{}##}$\hfil\crcr#1\crcr}}\,}
\newif\ifdtup

\def\Nc{$N_c\ $}
\def\Nf{$N_f\ $}

\def\Amuu{\vec{A}^\mu}
\def\Amud{\vec{A}_\mu}
\def\Anuu{\vec{A}^\nu}
\def\Anud{\vec{A}_\nu}
\def\Alud{\vec{A}_\lambda}
\def\Asud{\vec{A}_\sigma}

\def\dmuu{{\partial^\mu}}

\def\Ecas{$E_{cas}\ $}
\def\intif{\int_0^\infty}
\def\intifm{\int_M^\infty}
\def\undemi{ {1\over 2}}
\def\undemipet{ {\scriptstyle1\over\scriptstyle2}}
\def\dmuphi{\partial_\mu\phi}
\def\tr{{\rm Tr}}

\def\Lag{{\cal L}}

\def\X{{$\times$}}
\def\facpi{{96\pi^2}}
\def\xlog{\ln {M^2\over \mu^2} }

\def\pr{Phys. Rev. }

\def\np{Nucl. Phys. }
\begin{document}
\normalsize\textlineskip
\setcounter{page}{1}

\renewcommand{\thefootnote}{\fnsymbol{footnote}} 
\def\bsc{{\sc a\kern-6.4pt\sc a\kern-6.4pt\sc a}}
\def\bflatex{\bf L\kern-.30em\raise.3ex\hbox{\bsc}\kern-.14em
T\kern-.1667em\lower.7ex\hbox{E}\kern-.125em X}

\fpage{1}
\centerline{\bf CASIMIR ENERGY IN THE SKYRME MODEL\footnote{
Talk given at the workshop on "Baryons as Skyrme solitons", Siegen 27-30
September 1992} }
\vspace*{0.035truein}
\vspace{0.37truein}
\centerline{\footnotesize B. MOUSSALLAM}
\vspace*{0.015truein}
\centerline{\footnotesize\it Division de physique
th\'eorique\footnote{Unit\'e de recherche des universit\'es Paris 11
                     et Paris 6 associ\'ee au CNRS}
, Institut de physique nucl\'eaire}
\baselineskip=12pt
\centerline{\footnotesize\it  Universit\'e Paris-Sud, 91406 Orsay, France }
\vspace*{0.21truein}
\abstracts{
The so called Casimir energy embodies the O($(N_c)^0$) contribution
to the skyrmion mass according to the semi-classical expansion theory
of solitons. We claim that this contribution can be accurately estimated
despite the fact that some of the counterterms,
provided in principle by chiral perturbation theory, are
not known in practice. Using $\zeta$
function techniques we show that $E_{cas}=E_{cas}(\mu)+E_{ct}(\mu)$ where
$E_{cas}(\mu)>>E_{ct}(\mu)$ because it incorporates all the zero-mode
contributions and it can be exactly calculated. Our results confirm that
the fourth order Skyrme lagrangian does not seem to provide a correct
description of the lightest baryons as solitons. We show that a simple
extension to order six gives, on the contrary, good results without
tuning the parameters of the chiral lagrangian.
}{}{}

\textlineskip
\section{Introduction}

It is a common experience to most skyrmion practitioners that the  mass
tends to come out too high. Even if one uses  rather sophisticated
chiral lagrangians, incorporated vector and axial vector mesons (e.g.
\cite{vector}) one finds the predicted nucleon mass to lie nearly 50\%
too high and the situation becomes much worse in SU(3) extensions. Most
other observables, in contrast to that, seem to be predicted with a much
better accuracy. This has led to the belief that  the mass may have a
somewhat special status.

In this talk, I would like to try to convince you that this is not the case
and that the problem with the mass has a very natural explanation once
one tries to perform the semi-classical (which is also the 1/\Nc) expansion
in a systematic way. This, in fact, is not usually done in the context
of the skyrmion. Following ref.\cite{anw}, one identifies collective
coordinates and quantization is effected only at  the level of these
coordinates. If one does this for the mass, one finds besides the classical
O(\Nc) contribution a O(1/\Nc) one. This procedure gives no contribution
of order $N_c^0$. This contribution is precisely what one calls the
"Casimir energy" (CE) and its evaluation requires that one deals with the
non-collective coordinates.

\vfill

\noindent{IPNO/TH 92-94}
\eject

What one has to do in principle can be inferred
from the semi-classical soliton theory which was developped a long time
ago (see e.g.\cite{coleman} ). In practice, however, one seems to face
the difficulty that in early work, solitons were taken to live in 1+1
dimensions with a renormalizable  lagrangian, while the skyrmion lives
in 3+1 dimensions and the lagrangian is not renormalizable.
A safe approach to this problem is to use a lagrangian involving
only chiral field degrees of freedom. This is after all what Skyrme
originally did and we nowadays have a theory, Chiral perturbation theory
(ChPT)\cite{w76}\cite{gl84}, which precisely gives us such a lagrangian
and furthermore tells us how to make sense out of loops. Quite remarkably,
recent determinations of the values of the parameters appearing
in the lagrangian of ChPT turn out to be surprinsingly close to those
which Skyrme has guessed thirty years ago.

\section{Basic ideas and formulae}

Vibrational degrees of freedom for the skyrmion were first
considered in ref.\cite{biedenharn} who did not attempt, however, a
complete evaluation of their role. One approximation scheme to actually
calculate the CE was proposed by Schnitzer\cite{schnitzer} who
found a large positive value (of the order of 500 MeV). Later on, the presence
of vector mesons  in the lagrangian were found (in the same approximation)
to considerably reduce this effect to a negligible 50 MeV\cite{chemtob}.
Concurrently, an estimate was made in ref.\cite{zahed} which makes use of
the relationship between the CE and the effective action to one-loop. They
recognized that an ultra-violet divergence was present and they used a
derivative expansion approximation of the effective action. A somewhat better
approximation was proposed later on \cite{dobado} which seems nevertheless to
yield a rather similar result of approximately $-200$ MeV.

How come that all these results are different from each other? The answer is
that one must be very careful with the approximations that one makes. Most
approximations will in fact kill the most important part of the effect. To
show this, we will start from a formula which is exact and which relates the CE
to the pion-skyrmion phase shifts. The correct size of the effect will be
controlled by the fact that the phase-shifts are very large at the origin
because there a six zero-modes (associated with
three translations and three rotations). Any
approximation of the Born type for the phase-shifts will be essentially
incorrect. An approximation which respects the low-energy behaviour
of the phases was proposed in ref.\cite{nous}.

Let us start with a simple 1+1  dimensional situation as a warm up. Many
features are exactly similar to the 3 dimensional case.
Consider, for instance, the action:
$$
S=\undemi\int d^2x \left((\dmuphi)^2 +m^2\phi^2 -{\lambda\over 2}
\phi^4\right)
\eqno(1)$$
It is convenient to rescale the coordiates and the fields:
$$
\phi\to{m\over \sqrt{\lambda}}\phi\qquad x\to{1\over  m}x
$$
so the action now looks like:
$$
S={m^3\over 2\lambda}
\int d^2x \left( (\dmuphi)^2 +\phi^2 -\undemipet\phi^4\right)
\eqno(2)$$
We see that $1/\lambda$ appears in front of the action so the semi-classical
expansion is identical to the weak coupling expansion. We have exactly the
same situation in three dimensions with \Nc replacing $1/\lambda$. The
classical solution (the "kink") and the classical mass are easily found:
$$
\phi_{class}=\tanh({x-x_0\over \sqrt{2} }),\qquad M_{class}=
{2\sqrt{2}m^3\over 3\lambda}
\eqno(3)$$
Next, one has to consider fluctuations around the classical solution and it
is not difficult to show that the leading correction to the classical mass
(i.e. of order $\lambda^0$) has the following formal expression
\cite{coleman}:
$$
M^{(0)} =\undemi\left[
\tr\,\big(-\partial_x^2+3\tanh^2({x\over\sqrt{2}})-1\big)^\undemi-
\tr\,\big(-\partial_x^2+2                           \big)^\undemi\right]
\equiv\undemi\big(\tr H^\undemi -\tr H_0^\undemi\big)
\eqno(4)$$
Obviously if we rewrite the trace in terms of eigenvalues we obtain that
$$
M^{(0)} =\sum\omega_n-\sum\omega_n^0\eqno(5)$$
which is why this contribution
is called the Casimir energy, by analogy with the classic QED effect
\cite{casimir}. The operator involved in (4) is obtained by
expanding the action to second order in powers of the fluctuation. The
corresponding operator for the skyrmion is obtained by exactly the same
procedure. A careful derivation can be found in ref.\cite{moi}.

As was  mentioned
already, an extremely useful formula for practical purposes arises upon
expressing (4) in terms of phase shifts\cite{dhn}.
A simple direct derivation, valid for any space dimension,
is as follows. One first uses the identities:
$$
\tr H^\undemi=2\tr\intifm dE\, E^2\delta(E^2-H)=
-{1\over  i\pi}\tr\intifm dE\,E \left({E\over  E^2-H+i\epsilon}
                             -{E\over  E^2-H-i\epsilon}\right)
\eqno(6)$$
where $M$ is the lower bound of the continuous spectrum (in the kink
example $M=\sqrt{2}m$ ).
Next one recognizes that:
$$
\tr\left({E\over  E^2-H+i\epsilon}-{E\over  E^2-H_0+i\epsilon}\right)=
{d\over  dE}\ln\Delta^+,\eqno(7)$$
where
$$
\Delta^+={\rm det}\left(1-{1\over  E^2-H_0+i\epsilon}(H-H_0)\right)
\eqno(8)$$
So we obtain a formula:
$$
M^{(0)} ={1\over  4i\pi}\intifm dE\, E {d\over  dE}\,(\ln\Delta^+ -\ln\Delta^-)
\eqno(9)$$
(This takes care of the continuous part of the spectrum. Obviously, one
must also add the discrete part in the trace if there is any).
Now $\Delta^+$ is nothing but the Fredholm determinant.
In one space dimension it is a well
know fact from scattering theory that its phase is the phase-shift
\cite{newton}:
$$
\Delta^\pm=\vert\Delta\vert \exp(\pm i\delta(E))
\eqno(10)$$
so our expression (4) becomes:
$$
M^{(0)} ={1\over2\pi}\intifm dE E\,\delta'(E)
\eqno(11)$$
In order to generalize to higher dimensions we must return to (9). If
$H-H_0$ is a radial potential for instance
one can again express the phase of the
Fredholm determinant in terms of the phase-shifts of the radial operators
$\delta(E)\equiv{\rm phase}(\Delta^+)=\sum(2j+1)\delta_j(E)$. The
generalization
to the skyrmion case is obvious: for every value of the grand-spin
quantum number, J, we have three eigen-phase-shifts\cite{karliner}
so we must sum over these before summing over J.

At this point it would seem that we are running into trouble. A look
at refs.\cite{karliner} reveals that (except for J=0) the
phase-shifts are linearly diverging functions of $E$ and one can show
that after performing the J sum things become worse: one ends up with
a cubic divergence. Even in one dimension, in fact, the integral (11)
is logarithmically divergent. In that case, however ,
it is enough to remember that at order one in $\hbar$
there is an extra contribution to the energy which comes from the one-loop
counterterm in the lagrangian. Once this is taken into account the
divergence disappears. We will see later how one can generalize this
mechanism to the skyrmion. To begin with, we must discuss the lagrangian.

\section{Chiral perturbation theory and the skyrmion:}

In the limit where the masses of the \Nf
light quarks are set to zero the
QCD lagrangian is invariant under the chiral SU(\Nf)\X SU(\Nf) group. This
invariance is spontaneously broken by the QCD vacuum down to SU(\Nf). The
spectrum thus consists of \Nf zero-mass goldstone bosons and there is a mass
gap of around $\Lambda=1$ GeV above which one finds meson resonances,
baryons, etc....ChPT is a systematic framework for describing
low energy phenomena (with typical energy E) as an expansion
in powers of E$/\Lambda$. In fact, it is a more general expansion which
involves
quark masses, external fields etc... but let us assume that these are
zero for the moment. It is convenient to encode the goldstone bosons
in a unitary matrix on which the chiral group operates linearly:
$U=\exp(i\vec\tau.\vec\pi)$ then the most general dynamics can be
expressed with the aid of a chiral lagrangian:
$$
\Lag(U)=\Lag_2 +\Lag_4 +\Lag_6 +...
\eqno(12)$$
where the subscripts denote the numbers of derivatives of the matrix
field $U$. As shown by Weinberg\cite{w76} if one  wants to expand
an amplitude to
order E$^n$ with $n>2$ one must compute loops. There are simple
counting rules, for example
that one loop made out of two $\Lag_2$ vertices contribute
at order 4, one loop with  one $\Lag_2$ vertex and one $\Lag_4$ contributes
at order 6, a two-loop amplitude with $\Lag_2$ vertices is also of
order 6,  etc...
The chiral lagrangian contains the counterterms
which render the loops finite.

If we are able to construct the skyrmion out of the chiral lagrangian (12)
then, at least in principle, one should be able to renormalize the
Casimir energy, which is a one-loop effect. Note that it is not clear
that the chiral expansion does apply for the soliton but, after all,
one might expect   the average
energy of a pion "inside" a skyrmion to be of the order of 200 MeV
since the skyrmion size is expected to be of the order of 1 fm. So
why shouldn't we try?

The
chiral lagrangian has so far been computed up to order four\cite{gl84}
and it reads (external sources being switched off):
$$
\Lag_{ChPT}={F^2\over 2}\Amuu.\Amud +{1\over \facpi}(\bar l_1-1+\xlog)
\left( \Amuu.\Amud\right)^2 +{2\over \facpi}(\bar l_2-1+\xlog)
\left( \Amuu.\Anud\right)^2\eqno(13)$$
where
$$
i\vec\tau.\Amuu=U^\dagger\dmuu U,\qquad M\simeq m_\pi,\qquad F\simeq f_\pi.
$$
We see that a scale $\mu$ appears in (13). This is because it incorporates
counterterms. If one uses a regularization prescription like dimensional
regularization, one gets rid of the 1/(n-4) pole and one is left with a
scale dependence. A similar but  more convenient scheme for our purposes
is the zeta function one\cite{mckeon}
as we will see later. In order to discuss solitons we
must perform an \Nc expansion so we might start by assuming:
$$
\Lag^{(1)}=\Lag_{ChPT} (\mu=m_\rho)
\eqno(14)$$
where the superscript designates the \Nc order. This is a meaningful
assumption provided subleading terms in \Nc are very small for this
particular scale value $\mu=m_\rho$.
This seems indeed to be the case as follows from the recent work of
ref.\cite{riggenbach}
who have analysed  Kl4 decays and showed that this provides
a quantitatively neat test for this suppression. Note that the fact
that the terms of order $N_c^0$  in the chiral lagrangian are found to
be small by no means implies that the correction of the same order
to the skyrmion mass should also be small. However, this will turn out to be
important for the accuracy of the CE determination, as we will see.

Let us now rewrite the fourth order lagrangian using notations which are
familiar in the skyrmion context:
$$
\Lag^{(1)}_4={1\over 4e^2}\,(\Amuu\wedge\Anuu).(\Amud\wedge\Anud)+
{\gamma\over 2e^2}(\Amuu.\Amud)^2
\eqno(15)$$
The first question that one might ask is whether the values of the
parameters $e$ (Skyrme parameter) and $\gamma$ that are determined from
ChPT are compatible with a stable soliton solution. Remember in this
context that the $\gamma$ term tends to destabilize the soliton. An upper
bound for stability was found to be $\gamma < 0.12 $\cite{truong}. If we
look at the 1984 values of Gasser and Leutwyler we find:
$$
e=6.8\pm 4.15\qquad\qquad \gamma=-0.06\pm 0.2
$$
These numbers are not very conclusive because of the large error bars. Let us
now consider the more recent determination of Riggenbach
et al.\cite{riggenbach}:
$$
e=7.1\pm2.30\qquad\qquad\gamma=0.03\pm0.03
$$
The error bars are considerably smaller and we see that ChPT now seems
perfectly compatible with a stable skyrmion. The results are  in fact
strickingly similar with the numbers proposed by Skyrme 30 years ago:
$$
e=6.28\phantom{\pm2.30}\qquad\qquad\gamma=0\phantom{\pm0.20}
$$
Unfortunately, chiral order four is not enough for a consistent description
of the skyrmion. Firstly it is not consistent with the chiral expansion
because the virial theorem implies that the contribution of $\Lag_2$
is identical to that of $\Lag_4$ instead of being much larger. Secondly it
it is not consistent with the large \Nc expansion either, because,
 as we will soon
discover,  the \Nc contribution turns out to be  nearly identical
to the $N_c^0$ one instead, again, of being significantly larger.

The situation is perhaps not desperate. In fact I claim that a mere
extension to chiral order six is sufficient to cure, apparently, all the
difficulties. We need however to have a model in order to make a guess for
the coefficients of the terms appearing in $\Lag_6$ since they have not
yet been worked out from ChPT. A reasonable starting point seems to be
provided by the observation\cite{ecker} that all of the 10 parameters
which appear in the chiral order four lagrangian can be saturated to
a very good approximation by the contribution of the low-lying vector,
axial vector, scalar and pseudo-scalar resonances. In particular since
the rho meson saturates the Skyrme term it is natural to consider the omega
meson as well which contributes at sixth order:
$$
\Lag_{6,\omega}=-\undemi{\beta^2\over M^2_\omega}\, B_\mu B^\mu,\qquad
B_\mu ={\epsilon_
{\mu\nu\lambda\sigma}\over 12\pi^2}(\Anud,\Alud,\Asud)
\eqno(16)$$
The value of the coupling parameter $\beta$ can be estimated to be
rather large $\beta\simeq 9.3$\cite{vinhmau} so it is tempting
to assume that $\Lag_{6,\omega}$ is the dominant sixth order term. This
is supported by estimates by Walliser\cite{walliser} who showed that
the rho induced sixth-order term is much smaller than the omega one.
One also notices that the rho and scalar contributions tend to cancel
each other.

If one includes the contribution (16) in  the chiral lagrangian
and looks for a soliton solution, the result is found to be in much
better agreement with the chiral expansion than before.
Due to the repulsive effect
of the omega induced sixth order term the profile size is much larger
than before and, as a result, the contribution of $\Lag_2$ becomes
five times larger than that of $\Lag_4$. We will see in the sequel that
the \Nc expansion also seems to become coherent, but before we can actually
evaluate the CE in order to check that, we must discuss the
$\zeta$ function method of regularization which is a very important
technical ingredient in the calculation.

\vfill\eject
\section{$\zeta$ function regularization method}

The basic idea is to introduce instead of a sum like $\sum (\omega^2)^
\undemi$ the sum $\sum (\omega^2)^{\undemi-s}$ depending on the complex
parameter $s$. One can eventually show that the function is analytic in
$s$ and further define the limit of interest, $s=0$, by analytic continuation.
The non trivial part is to actually perform this analytic continuation in a
situation like ours where the eigenvalues $\omega_n$ are known only
numerically. As we will see, the phase-shift representation provides a
simple solution to this problem. Before we turn to that, however, we
must make sure that the regularization procedure that we use for the Casimir
energy is the same as the one which is used to regularize Green's functions
(leading to the counterterms as they appear in (13) ). Now Green's functions
to one loop are generated by an effective action which can be written as
a trace log of a four dimensional operator. If we call $O$ such an operator
then the corresponding $\zeta$ function is defined as\cite{ramond}:
$$
\zeta_O(s)={1\over\Gamma(s)}\tr\,\intif d\tau\,\tau^{s-1}
\exp\left(-{\tau O\over\mu^2}\right)
\eqno(17)$$
the scale $\mu$ is introduced at this level in order to make $\tau$
dimensionless. The regularized form of the trace log is then given by:
$$
\tr\log(O)\equiv-\zeta '(s=0)
\eqno(18)$$
we can relate the Casimir energy to a four dimensional operator via the
identity:
$$
\tr(H^\undemi)=\lim_{T\to\infty}{1\over T}\tr\log\, (-\partial_t^2 + H)
\eqno(19)$$
which holds for a time independent operator $H$. If we let $O=(-\partial_t^2
+ H)$ in formula (17) a simple calculation shows that the $\zeta$ function
regularization of the Casimir energy which matches the one used for the
effective action is:
$$
\tr(H^\undemi)\equiv -\zeta '(0)\quad{\rm with}\quad\zeta(s)=-{\mu^{2s}\Gamma
(s-\undemi)\over\Gamma(s)\Gamma(-\undemi)}\undemipet\tr H^{\undemi-s}
\eqno(20)$$
The same derivation as before yields the appropriate phase-shift formula:
$$
\tr H^{\undemi-s}-\tr H_0^{\undemi-s}={1\over\pi}\intif\,dp\,\delta '(p)
(p^2 +M^2)^{\undemi-s}
\eqno(21)$$
where a finite pion mass $M$ was introduced. It can be shown that the
large momentum behavior  of the phase function is of the form:
$$
\delta(p)=\bar a_0 p^3 + \bar a_1 p +{\bar a_2\over p} + ...
\eqno(22)$$
where the $\bar a_i$'s are numbers which are simply related to the heat kernel
expansion of the operator $H$. One can subtract and add this leading
asymptotic behavior and perform the analytic continuation in $s$ by
using the well-known integrals\cite{abramowitz}
$$
\intif dp p^m (p^2+M^2)^{\undemi-s}=\undemi M^{m+2-2s}{\Gamma({m+1\over2})
\Gamma(s-1-{m\over2})\over\Gamma(s-\undemi)}
\eqno(23)$$
Finally, one obtains the finite and closed form formula for the
regularized CE:
$$\eqalign{
E_{cas}(\mu)=
&{1\over2\pi}\bigg\lbrace \intif dp\,\Big[-{p\over\sqrt{p^2+M^2}}
(\delta(p)-\bar a_0p^3-\bar a_1p) + {\bar a_2\over\sqrt{p^2+\mu^2}}\Big]\cr
&-{3 \bar a_0\over8}M^4({3\over4}+{1\over2}\ln{\mu^2\over M^2})
+{\bar a_1\over4}M^2 (1+\ln{\mu^2\over M^2})-M\delta(0)\bigg\rbrace\cr
}\eqno(24)$$
Note that it is well defined for zero as well as finite pion mass $M$. It
is also clear that, by construction, only the low-energy behavior of the
phase shift is important in the integration.
Now, in a way similar to the 1+1 dimensional case we must add counterterms
to (24), i.e. the O($(N_c)^0$) part of the chiral lagrangian:
$$
E_{ct}(\mu)=- (\Lag_2^{(1)}+...+\Lag_n^{(1)}  - \Lag_{ChPT} )
\eqno(25)$$
Actually, since $\Lag^{(1)}$ is being truncated at chiral order $n$
(in practice $n=4$ or 6) $E_{ct}$ contains terms of order O(\Nc) and
of chiral order $n+2,..,2n$. Consistency requires that these should
be of the same order in magnitude (or smaller) that the terms of lower
chiral order but subleading in \Nc. In this respect already, $n=6$
is more satisfactory than $n=4$.
Next, when we add the two pieces:
$$
M^{(0)}=E_{cas}(\mu)+E_{ct}(\mu)
\eqno(26)$$
the scale dependence should disappear (at least up to O(1/\Nc) terms).
In practice, of course, it does not since we do not know enough terms
in $\Lag_{ChPT}$. One would therefore like to argue that
$$
E_{ct}(\mu)<< E_{cas} (\mu)
\eqno(27)$$
This, of course, cannot hold for arbitrary values of $\mu$. Now as
we have seen, it follows from ref.\cite{riggenbach} ( and
also from ref.\cite{ecker} ) that
for the particular value $\mu=m_\rho$ terms of order $N_c^0$ are
strongly suppressed in the chiral lagrangian (by a factor of 10 or so)
so it seems a good idea for us to choose $\mu=m_\rho$. According to
formula (25) this will suppress the contributions to $E_{ct}$ up to chiral
order $n$. Those of order $n+2, n+4,...$ are of order \Nc so one must
assume that they are suppressed because of they high chiral order. The
main reason why (27) should hold, though, is that $E_{cas}(\mu)$ is enhanced
because it incorporates the zero-mode contributions. In our formalism,
they show up via Levinson's theorem forcing the phase-shift at the
origin to be fairly large (=$6\pi$). An approximate way to estimate
the CE by singling out the zero-mode contributions was imagined recently
by Holzwarth\cite{holzwarth}.
Let us now illustrate these points on some examples.

\section{Some results and conclusions}

Let us first consider the Skyrme lagrangian $\Lag=\Lag_2+\Lag_{4,sk}$
(see (15) ) with physical values of $f_\pi$ and $m_\pi$ and

\noindent a)e=5.5 (by analogy with ref.\cite{anw}).
Following the method
described above one finds:
$$
M^{(0)}=-957 + (-72) =-1029\ {\rm MeV}
$$
where the number in parenthesis is the counterterm contribution
(which is indeed small). This is
to be compared with the leading \Nc contribution:
$$
M^{(1)}=1263\,{\rm MeV}
$$
Clearly, the 1/\Nc expansion seems to be in trouble and
furthermore, the nucleon mass is found
to be much too small. In fact, better phenomenological results are expected
from the Skyrme lagrangian if one takes a smaller value for e. Let us consider
then:

\noindent b)e=4.0 (which gives for example a reasonable delta-nucleon
splitting of 250 MeV). In this case, one finds:
$$
M^{(0)}=-805 + (-452) =-1297 {\rm MeV}
$$
while the leading contribution is
$$
M^{(1)}=1761 {\rm MeV}
$$
This is slightly better than before and one could eventually
arrive at a reasonable
nucleon mass, but an unsatisfactory feature now is
that the counterterm contribution is rather large. This is because of the
large mismatch between the value of $e$ and the one compatible with ChPT
(see sec. (3) ). In this situation one can no longer argue that the unknown
contributions of $E_{ct}$ are necessarily small.

The only way out of this
dilemma seems to include higher order terms in the chiral lagrangian. Let
us add an omega induced sixth order term then, and consider:$\Lag^{(1)}=
\Lag_2+\Lag_{4,sk}+\Lag_{6,\omega} $ with the parameters $e=7.22$ (from
the meson saturation fit of \cite{ecker}) and $\beta=9.3$ from
\cite{vinhmau}. The calculation
gives, in that case:
$$
M^{(0)} =- 604 + (+153)= -451 {\rm MeV}
$$
while the classical value
$$
M^{(1)} =1553 {\rm MeV}
$$
The stricking feature is the strong reduction of the Casimir contribution.
Now the large \Nc expansion looks much more reasonable than before (the
correction is a factor of 3 smaller than the classical mass). We
argued in sec. (3) that the chiral expansion was also more justified. The
two things are in fact related: the reason why \Ecas is smaller is that the
phase-shift function $\delta(p)$ drops faster as a function of the momentum,
and this is because the classical profile function has a larger
extension in space. Note that the counterterm contribution is positive now
(because the choice of $e$ is slightly larger than that of ChPT) but the
Casimir energy itself is always found to be negative. If we add $M^{(1)}$
and $M^{(0)}$
we find that the nucleon mass is correctly predicted to within 20\% while
we did not attempt to fit any parameter in the chiral lagrangian.

In conclusion, a  reasonable picture seems to emerge
provided one incorporates besides the fourth order rho induced term
a sixth order omega meson one. This is not really a surprise. In fact,
the most succesfull skyrmion phenomenology seems to require that one has
these resonances (and a few others) explicitely in the lagrangian. The
rationale for including resonances
is that this should extent the range of validity of the effective
lagrangian from $E<<\Lambda$ to $E\simeq\Lambda$. It is likely that one could
extend the calculation of the CE to that case but the value of the "optimal
scale" should perhaps be larger. An open question at the moment is how to
estimate loop corrections for other nucleon observables. In particular, it
is not clear (to me) whether the zero-mode enhancement which lead to a
sizeable effect in the case of the mass will also operate for some other
observables and how.

\nonumsection{References}


\begin{thebibliography}{99}
\bibitem{vector}N. Kaiser, these proceedings.
\bibitem{anw}G.S. Adkins, C.R. Nappi and E. Witten, \np B228, 552 (1983)
\bibitem{coleman}S. Coleman, "Aspects of symmetry", Cambridge
University Press (1985) Chap. 6
\bibitem{w76}S. Weinberg, Physica A96, 327 (1979)
\bibitem{gl84}J. Gasser and H. Leutwyler, Ann. Phys. 158, 142 (1984)
\bibitem{biedenharn}L.C. Biedenharn, Y. Dothan and M. Tarlini,
\pr D31, 649 (1985)
\bibitem{schnitzer}H. Schnitzer, Nucl. Phys. B261, 546 (1985)
\bibitem{chemtob}M. Chemtob, \np A473, 613 (1987)
\bibitem{zahed}I. Zahed, A. Wirzba and U-G. Meissner, Phys. Rev. D33 (1986)
830
\bibitem{dobado}A. Dobado and J. Terron, Phys. Lett. B247 (1990) 581
\bibitem{nous}B. Moussallam and D. Kalafatis, Phys. Lett. B272, 196 (1991)
\bibitem{casimir}C. Itzykson and J-B Zuber, "Quantum field theory",
Mc Graw-Hill (1980), p.138.
\bibitem{moi}B. Moussallam, preprint IPNO/TH 92-75, submitted to
Annals of Physics.
\bibitem{dhn}R.F. Dashen, B. Hasslacher and A. Neveu, \pr D10, 4114,
4130  (1974)
\bibitem{newton}R.G. Newton, "Scattering theory of waves and particles",
Mc Graw-Hill (1982)
\bibitem{karliner}H. Walliser and G. Eckart, Nucl. Phys. A429, 514 (1984) ,
M.P. Mattis and M. Karliner, Phys. Rev. D31, 2833 (1985)
\bibitem{mckeon}D.G.C. McKeon and T.N. Sherry, Phys. Rev. D35, 3854 (1987)
\bibitem{riggenbach}C. Riggenbach, J. Gasser, J.F. Donoghue and B.R. Holstein,
\pr D43, 127 (1991)
\bibitem{truong}T.N. Pham and T.N. Truong, \pr d31, 3027 (1985)
\bibitem{ecker}G. Ecker, J. Gasser, A. Pich, E. de Rafael, \np B321, 311
(1989)
\bibitem{vinhmau}M. Lacombe, B. Loiseau, R. Vinh Mau and W.N. Cottingham,
\pr D38, 1491 (1988)
\bibitem{walliser}H. Walliser, these proceedings.
\bibitem{ramond}P. Ramond, "Field theory: a modern primer", Addison-Wesley,
New York (1981)
\bibitem{abramowitz}"Handbook of mathematical functions", ed. M. Abramowitz
and I.A. Stegun, Dover pub., p258.
\bibitem{holzwarth}G. Holzwarth, these proceedings.
\end{thebibliography}
\end{document}